\documentclass[12pt]{article}
\usepackage{amssymb}
\usepackage{graphicx}
\usepackage{subfigure}
\usepackage{caption2}

\begin{document}

\title{Stability and collisions of moving semi-gap solitons in Bragg
cross-gratings\thanks{
\textit{Preprint submitted to Physics Letters A}}}
\author{T. Mayteevarunyoo$^{1}$, Boris A. Malomed$^{2}$, P.L. Chu$^{3}$, and
A. Roeksabutr$^{1}$ \\
{\small $^{1}$\textit{Department of Telecommunication Engineering,
Mahanakorn University}}\\
{\small \textit{of Technology, Bangkok10530, Thailand}}\\
\textit{\ }{\small $^{2}$\textit{Department of Interdisciplinary Studies,
Faculty of Engineering,}}\\
\textit{\ {\small Tel Aviv University, Tel Aviv 69978, Israel}}\\
{\small $^{3}$\textit{Optoelectronics Research Centre, Department of
Electronics}}\\
\textit{\ {\small Engineering, City University of HongKong}}}
\maketitle

\begin{abstract}
We report results of a systematic study of one-dimensional four-wave moving
solitons in a recently proposed model of the Bragg cross-grating in planar
optical waveguides with the Kerr nonlinearity; the same model applies to a
fiber Bragg grating (BG)\ carrying two polarizations of light. We
concentrate on the case when the system's spectrum contains no true bandgap,
but only semi-gaps (which are gaps only with respect to one branch of the
dispersion relation), that nevertheless support soliton families. Solely
zero-velocity solitons were previously studied in this system, while current
experiments cannot generate solitons with the velocity smaller than half the
maximum group velocity. We find the semi-gaps for the moving solitons in an
analytical form, and demonstrated that they are completely filled with
(numerically found) solitons. Stability of the moving solitons is identified
in direct simulations. The stability region strongly depends on the
frustration parameter, which controls the difference of the present system
from the usual model for the single BG. A completely new situation is
possible, when the velocity interval for stable solitons is limited not only
from above, but also from below. Collisions between stable solitons may be
both elastic and strongly inelastic. Close to their instability border, the
solitons collide elastically only if their velocities $c_{1}$ and $c_{2}$
are small; however, collisions between more robust solitons are elastic in a
strip around $c_{1}=-c_{2}$.
\end{abstract}

\section{Introduction}

Bragg gratings (BGs) are broadly used as a basis for spectral
filters are other elements of optical telecommunication networks
\cite{Kashyap}. Simultaneously, BGs offer a unique medium for the
theoretical \cite{AcevesWab} -- \cite{defect} and experimental
\cite{experiment} study of \textit{gap solitons} (GSs), supported
by the balance between the BG-induced dispersion and Kerr (cubic)
nonlinearity dispersion . The latter topic was surveyed in an
earlier review \cite{OldReview} and recent one \cite{NewReview}.
Matter-wave pulses similar to the optical BG solitons were very
recently observed in Bose-Einstein condensates \cite{BEC}.

The BG solitons may exist not only as temporal pulses in fiber
gratings, but also as spatial solitons in 2D (two-dimensional)
waveguides \cite{Feng}. In the latter case, the BG is represented
by a set of parallel grooves on the surface of the waveguide. A
generalization, that makes it possible to increase the number of
waves coupled by the BG, is based on a superposition of more than
one BGs in the spatial domain. In particular, a system of three
gratings which form a triangular lattice was considered in Ref.
\cite{Georg}. It couples three waves, giving rise to new types of
stable GSs -- for instance, ``peakons".

In recent papers \cite{IlyaTemporal} and \cite{IlyaSpatial}, a model of a
\textit{cross-grating} was introduced, as a superposition of two mutually
transverse BGs written on a 2D waveguide with the Kerr nonlinearity. In this
case, one is dealing with a system of four waves. The evolution of their
amplitudes $v_{j}$, $j=1,2,3,4$, obeys the system of normalized equations
\cite{IlyaTemporal,IlyaSpatial},
\begin{eqnarray}
{\small i}\frac{\partial v_{1}}{\partial
t}+\frac{i}{\sqrt{2}}\frac{\partial v_{1}}{\partial
z}+\frac{i}{\sqrt{2}}\frac{\partial v_{1}}{\partial x}+v_{2}+\mu
v_{3}+\left[ \left\vert v_{1}\right\vert ^{2}+2(\left\vert
v_{2}\right\vert ^{2}+\left\vert v_{3}\right\vert ^{2}+\left\vert
v_{4}\right\vert ^{2})\right] v_{1}+2v_{2}v_{3}v_{4}^{\ast }
&=&{\small 0,}
\nonumber \\
i\frac{\partial v_{2}}{\partial
t}-\frac{i}{\sqrt{2}}\frac{\partial v_{2}}{\partial
z}+\frac{i}{\sqrt{2}}\frac{\partial v_{2}}{\partial x}+v_{1}+\mu
v_{4}+\left[ \left\vert v_{2}\right\vert ^{2}+2(\left\vert
v_{1}\right\vert
^{2}+\left\vert v_{3}\right\vert ^{2}+\left\vert v_{4}\right\vert ^{2})\right] v_{2}+2v_{1}v_{4}v_{3}^{\ast } &=&0,  \nonumber \\
i\frac{\partial v_{3}}{\partial
t}+\frac{i}{\sqrt{2}}\frac{\partial v_{3}}{\partial
z}-\frac{i}{\sqrt{2}}\frac{\partial v_{3}}{\partial x}+v_{4}+\mu
v_{1}+\left[ \left\vert v_{3}\right\vert ^{2}+2(\left\vert
v_{1}\right\vert
^{2}+\left\vert v_{2}\right\vert ^{2}+\left\vert v_{4}\right\vert ^{2})\right] v_{3}+2v_{1}v_{4}v_{2}^{\ast } &=&0,  \nonumber \\
i\frac{\partial v_{4}}{\partial
t}-\frac{i}{\sqrt{2}}\frac{\partial v_{4}}{\partial
z}-\frac{i}{\sqrt{2}}\frac{\partial v_{4}}{\partial x}+v_{3}+\mu
v_{2}+\left[ \left\vert v_{4}\right\vert ^{2}+2(\left\vert
v_{1}\right\vert ^{2}+\left\vert v_{3}\right\vert ^{2}+\left\vert
v_{2}\right\vert ^{2})\right] v_{4}+2v_{2}v_{3}v_{1}^{\ast } &=&0.
\label{tzx}
\end{eqnarray}Here, $t$ is time, $x$ and $z$ are the coordinates in the waveguide, the
Bragg reflectivity (strength) of one grating is normalized to be
$1$, and $\mu $ is a relative strength of the second grating.
Coefficients in front of the nonlinear terms in Eqs. (\ref{tzx}),
which represent the self-phase-modulation, cross-phase-modulation
and four-wave-mixing, correspond to the normal Kerr effect in
isotropic materials.

The same model finds a different interpretation in terms of a 3D
photonic crystal with a transverse chess-board structure
\cite{IlyaTemporal,IlyaSpatial}. In that case, Eqs. (\ref{tzx}),
with $t$ replaced by the third coordinate $y$, along which the
medium is homogeneous, govern the spatial evolution of the
time-independent fields. In the latter context, a model which is
tantamount to the particular symmetric case of the system
(\ref{tzx}), with $\mu =1$, was earlier introduced in Ref.
\cite{John} (in fact, this case is a degenerate one, as it
corresponds to the gap in the model's linear spectrum collapsing
to a point \cite{IlyaTemporal,IlyaSpatial}). Ref. \cite{John} was
dealing with 2D solitons, treating their stability by means of a
method which is tantamount to the known Vakhitov-Kolokolov (VK)
criterion \cite{VK}. In fact, this method is not sufficient for
the verification of the stability, and our direct simulations (to
be reported elsewhere) demonstrate that \emph{all} the 2D
solitons, both in the general system (\ref{tzx}) and in its
special case with $\mu =1$ that was considered in Ref.
\cite{John}, are \emph{unstable}.

A simpler possibility is to consider 1D solitons, which are generated by the
substitution $v_{j}(t,x,z)\equiv e^{ikz}u_{j}(t,x)$ in Eqs. (\ref{tzx}),
\begin{eqnarray}
i\frac{\partial u_{1}}{\partial
t}-\frac{k}{\sqrt{2}}u_{1}+\frac{i}{\sqrt{2}}\frac{\partial
u_{1}}{\partial x}+u_{2}+\mu u_{3}+\left[ \left\vert
u_{1}^{2}\right\vert +2(\left\vert u_{2}^{2}\right\vert
+\left\vert u_{3}^{2}\right\vert +\left\vert u_{4}^{2}\right\vert
)\right]
u_{1}+2u_{2}u_{3}u_{4}^{\ast } &=&0,  \nonumber \\
i\frac{\partial u_{2}}{\partial
t}+\frac{k}{\sqrt{2}}u_{2}+\frac{i}{\sqrt{2}}\frac{\partial
u_{2}}{\partial x}+u_{1}+\mu u_{4}+\left[ \left\vert
u_{2}^{2}\right\vert +2(\left\vert u_{1}^{2}\right\vert
+\left\vert u_{3}^{2}\right\vert +\left\vert u_{4}^{2}\right\vert
)\right]
u_{2}+2u_{1}u_{4}u_{3}^{\ast } &=&0,  \nonumber \\
i\frac{\partial u_{3}}{\partial
t}-\frac{k}{\sqrt{2}}u_{3}-\frac{i}{\sqrt{2}}\frac{\partial
u_{3}}{\partial x}+u_{4}+\mu u_{1}+\left[ \left\vert
u_{3}^{2}\right\vert +2(\left\vert u_{1}^{2}\right\vert
+\left\vert u_{2}^{2}\right\vert +\left\vert u_{4}^{2}\right\vert
)\right]
u_{3}+2u_{1}u_{4}u_{2}^{\ast } &=&0,  \nonumber \\
i\frac{\partial u_{4}}{\partial
t}+\frac{k}{\sqrt{2}}u_{4}-\frac{i}{\sqrt{2}}\frac{\partial
u_{4}}{\partial x}+u_{3}+\mu u_{2}+\left[ \left\vert
u_{4}^{2}\right\vert +2(\left\vert u_{1}^{2}\right\vert
+\left\vert u_{3}^{2}\right\vert +\left\vert u_{2}^{2}\right\vert
)\right] u_{4}+2u_{2}u_{3}u_{1}^{\ast } &=&0.  \label{tx}
\end{eqnarray}Note that Eqs. (\ref{tx}) conserve the Hamiltonian, momentum, and the norm
(which is frequently called energy, in the applications to optics),
\begin{equation}
E=\sum_{j=1}^{4}\int_{-\infty }^{+\infty }\left\vert u_{j}(x)\right\vert
^{2}dx.  \label{E}
\end{equation}

The model based on Eqs. (\ref{tx}) is of special interest, as
previously considered multi-wave GS models, such as the three-wave
one in Ref. \cite{Mak}, did not contain a parameter similar to
$k$, which accounts for \textit{frustration}, i.e., symmetry
breaking in the wave pairs $\left( u_{1},u_{2}\right) $ and
$\left( u_{3},u_{4}\right) $ in Eqs. (\ref{tx}). In fact, the
frustration makes the present four-wave distinct from the usual
two-wave BG model.

The same system (\ref{tx}) also finds an interpretation in terms
of the BG in optical fibers. In that case, the fields $u_{1,2}$
and $u_{3,4}$ in Eqs. (\ref{tx}) represent pairs of right- and
left-travelling waves with two orthogonal circular polarizations,
$\mu $ is the Bragg-reflectivity strength, while the
linear-coupling terms with the coefficient $1$ account for linear
mixing between the circular polarizations due to an elliptic
deformation of the fiber's core, and the terms proportional to $k$
take into regard birefringence due to a twist of the fiber.

Looking for solutions to the linearized version of Eqs. (\ref{tx})
as $u_{j}(x,t)\sim \exp \left( i\lambda x-i\omega t\right) $, one
arrives at a dispersion relation,\begin{equation} \omega
^{2}=\frac{1}{2}\left( \sqrt{2\mu ^{2}+\lambda ^{2}}\pm
\sqrt{2+k^{2}}\right) ^{2},  \label{omega^2}
\end{equation}which generates a bandgap in the spectrum, $\left\vert \omega \right\vert
<\omega _{\mathrm{edge}}\equiv \mu -\sqrt{1+\left( k^{2}/2\right)
}$, provided that $\mu ^{2}>1+\left( k^{2}/2\right) $. Besides the
gap, the two branches of the dispersion relation (\ref{omega^2})
give rise to \textit{semi-gaps}, $\max \left\{ \mu -\sqrt{1+\left(
k^{2}/2\right) },0\right\} <\left\vert \omega \right\vert <\mu
+\sqrt{1+\left( k^{2}/2\right) }$, which are also found in the
case of
\begin{equation}
\mu ^{2}<1+\left( k^{2}/2\right) ,  \label{no-gap}
\end{equation}when the full gap is absent. It was recently shown, in a general form, that
families of BG solitons may exist in semi-gaps even if the full gap is
absent \cite{Mak}.

Quiescent (zero-velocity) solitons in Eqs. (\ref{tx}), of the form
$u_{j}(x,t)=U_{j}(x)\exp \left( -i\omega t\right) $, were
investigated in detail in Ref. \cite{IlyaTemporal} [solitons
generated by Eqs. (\ref{tzx}) in the spatial domain, corresponding
to $u_{j}(t,z,x)=\exp \left( -i\omega t\right) U_{j}(z,x)$ with
fixed $\omega $, were studied in Ref. \cite{IlyaSpatial}]. They
obey a reduction $U_{3}=-U_{1}^{\ast },\,U_{4}=-U_{2}^{\ast }$,
which is not valid for moving solitons, see below. If $k=0$, the
solitons fall into three categories: symmetric and anti-symmetric
ones, with $U_{1}=\pm U_{2}$, and more general asymmetric
solitons. The latter ones, and the solitons obtained by their
continuation to $k\neq 0$, are always subject to a weak
instability. The symmetric and anti-symmetric solitons, as well as
those stemming from them at $k\neq 0$, were found, respectively,
to be stable chiefly (but not exclusively) in the full gap (if
any), and in the semi-gaps outside the full gap.

Moving solitons are to be looked for solitons in the form
\begin{equation}
u_{j}(x,t)=\exp \left( -i\omega t\right) U_{j}(\xi ),~\xi \equiv x-ct.
\label{c}
\end{equation}Straightforward consideration of the underlying system (\ref{tx}) shows that
moving solitons may exist for $|c|~<c_{\max }=1/\sqrt{2}$ \cite{IlyaTemporal}
. On the other hand, quiescent BG solitons have never been observed in the
experiment, the minimum velocity at which they could be created being,
roughly, $c_{\max }/2$ \cite{experiment} (a possibility to diminish the
velocity through collisions between the moving solitons was proposed in Ref.
\cite{Mak-moving}). In Ref. \cite{IlyaTemporal}, moving solitons in the
present model were considered, in a brief form, inside the full gap. It was
found that they loose their stability at quite small values of the velocity,
$\simeq 0.2c_{\max }$. Our aim is to study the stability of the moving
solitons in a systematic way, as well as collisions between them. We will
consider moving solitons in the most interesting case, when the full bandgap
does not exist, but, nevertheless, solitons may be found in the semi-gaps
(see above). To the best of our knowledge, the moving solitons in this case
has never been considered before.

\section{Existence and stability of the moving solitons}

\begin{figure}[tbp]
\renewcommand{\captionfont}{\small \sffamily} \renewcommand{\captionlabelfont}{} \centering\subfigure[]{\label{fig:subfig:a}
\includegraphics[width=3.2in]{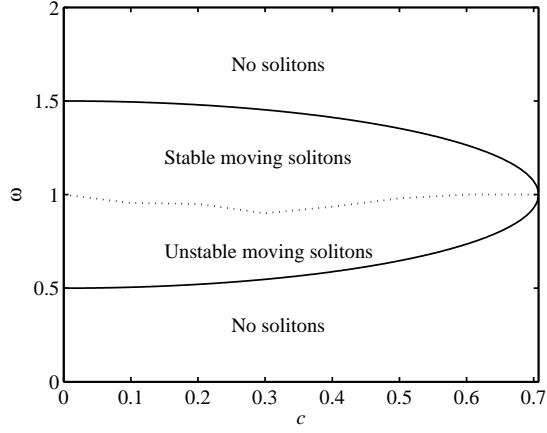}}
\subfigure[]{\label{fig:subfig:b}
\includegraphics[width=3.2in]{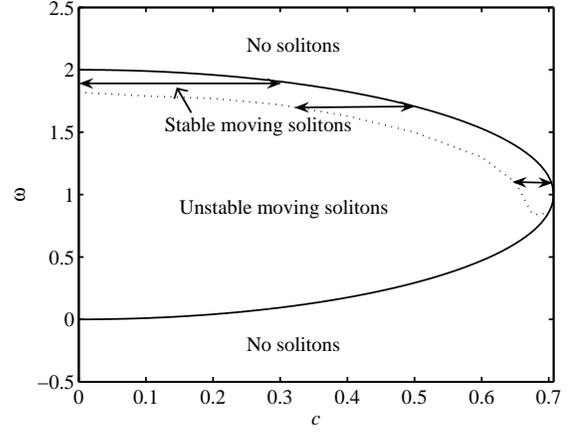}}
\subfigure[]{\label{fig:subfig:c}
\includegraphics[width=3.2in]{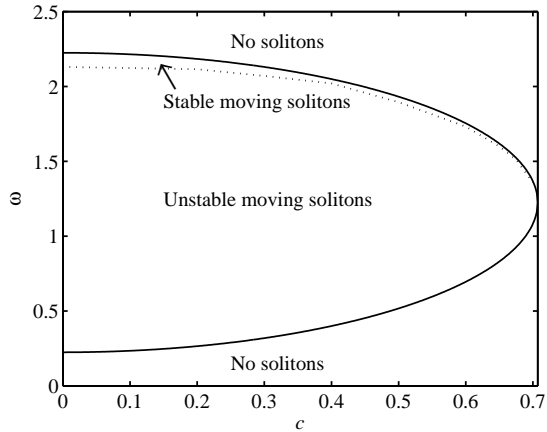}}
\par
\renewcommand{\figurename}{Fig.}
\caption{The solid curves are edges of the semi-gap, determined in
the analytical form by Eq. (\protect\ref{edge}) with the lower
sign in the left-hand side [a counterpart of this semi-gap,
corresponding to the upper sign in Eq. (\protect\ref{edge}), is
located at $\protect\omega <0$]. The region between the solid
curves (the semi-gap proper) is completely filled by the solitons,
the dotted curve being the border between stable and unstable
ones, as identified in direct simulations. Shown are typical
examples for the cases of zero frustration, $k=0$,$\protect\mu
=0.5$ (a), small frustration, $k=0.005$,$\protect\mu =1$ (b), and
large frustration, $k=1$,$\protect\mu =1$ (c). Three double arrow
lines in (b) are cross sections of the stability region for
$\protect\omega =\mathrm{const}$, which correspond to Eqs.
(\protect\ref{minmax}).} \label{semigaps}
\end{figure}

To look for solitons in the form (\ref{c}), one can rewrite the
underlying equations (\ref{tx}) in terms of the variables $(t,\xi
)$. The linearization of the equations gives rise to a dispersion
relation, which differs from (\ref{omega^2}) by the replacement
$\omega \rightarrow \omega +c\lambda $. In the case when the
spectrum contains no full gap, straightforward analysis yields an
equation to for the values of $\omega $ at edges of the semi-gaps,
where we aim to find the solitons:\begin{equation} \left[ \omega
\pm \sqrt{1+\left( k^{2}/2\right) }\right] ^{2}=\mu ^{2}\left(
1-2c^{2}\right) ~.  \label{edge}
\end{equation}It immediately follows from Eq. (\ref{edge}) that no semi-gaps, hence no
solitons either, are possible for $|c|~>1/\sqrt{2}$. Typical
examples of the semi-gap in the plane $\left( c,\omega \right) $
are shown in Fig. \ref{semigaps}.

Stationary equations, which are obtained by the substitution of
Eq. (\ref{c}) into Eqs. (\ref{tx}), were solved by means of the
shooting method. The numerical solution shows that the semi-gaps
are, indeed, completely filled by solitons. Note that the moving
solitons cannot be generated automatically from quiescent ones, as
Eqs. (\ref{tx}) bear no Galilean or Lorentzian invariance.

\begin{figure}[tbp]
\renewcommand{\captionfont}{\small \sffamily} \renewcommand{\captionlabelfont}{} \centering
\subfigure[]{\label{fig:subfig:a}
\includegraphics[width=4in]{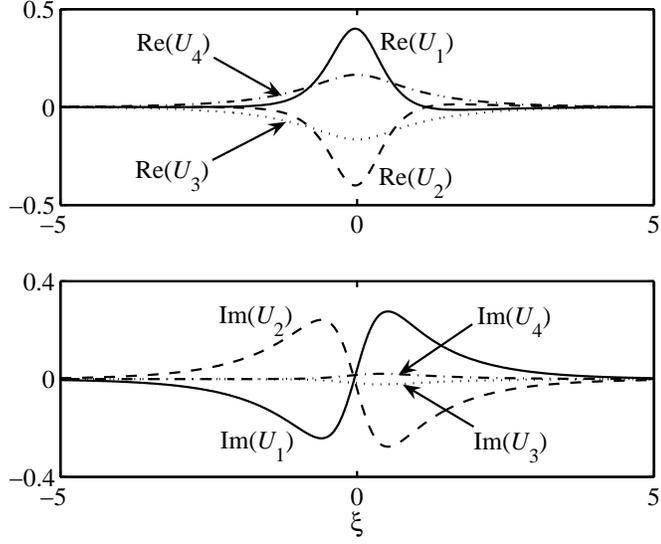}}
\subfigure[]{\label{fig:subfig:b}
\includegraphics[width=4in]{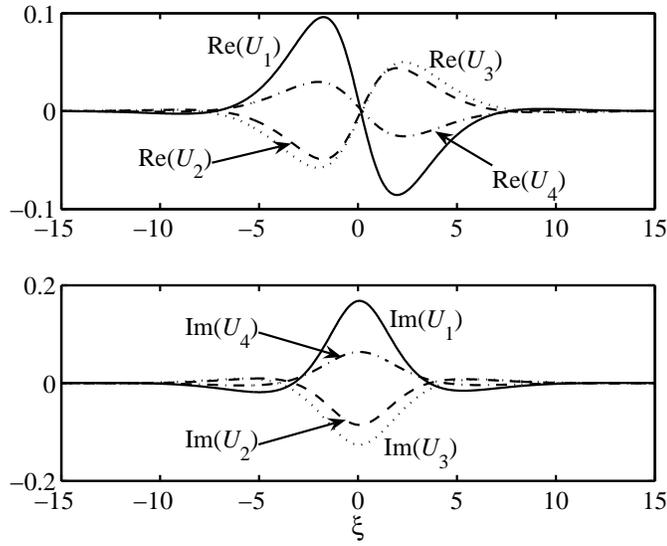}} \renewcommand{\figurename}{Fig.}
\caption{Generic examples of stable moving solitons in the form of
Eq. (\protect\ref{c}), found by the shooting method. Shown are
real and imaginary parts of the four fields. The panel (a) and (b)
represent, respectively, the cases of zero ($k=0,\protect\mu
=0.5,c=0.5,\protect\omega =1$) and finite ($k=1,\protect\mu
=1,c=0.2,\protect\omega =2.13$) frustration.} \label{solitons}
\end{figure}

Stability of the solitons was identified in direct simulations of
Eqs. (\ref{tx}) for perturbed solutions, by means of the
split-step Fourier-transform and finite-difference techniques,
both methods yielding identical results. Typical examples of the
stable moving solitons, found inside the semi-gaps, are displayed
in Fig. \ref{solitons}.

As well as in the case of zero-velocity solitons \cite{IlyaTemporal}, the
frustration parameter $k$ strongly affects the solitons' shape and
stability: if $k=0$, the stability area is largest, and the stable solitons
obey the reduction $U_{1,3}=-U_{2,4}$. As is seen in Figs. \ref{solitons}
and \ref{semigaps}, nonzero frustration breaks the latter relations, and
leads to strong shrinkage of the stability region.

A notable feature is that, in the case of small but finite $k$,
the stability interval for a fixed value of $\omega $ may be
limited not only by a maximum value of the soliton's velocity, but
also by a \emph{minimum} one. For example, three cross sections
shown by arrows in Fig. \ref{semigaps}(b) (for $k=0.005$ and $\mu
=1$)  correspond to the stability intervals\begin{eqnarray}
0 &\leq &|c|<0.3082~\mathrm{for}~\omega =1.9;~  \nonumber \\
0.28 &<&|c|<0.5049~\mathrm{for}~\omega
=1.7,~0.6595<|c|<0.70356~\mathrm{for}~\omega =1.1,  \label{minmax}
\end{eqnarray}the two latter ones being limited by a minimum value of $c$. The fact that
the stability of the solitons may require a finite velocity is a completely
new one. Nothing similar occurs in the model supporting the usual two-wave
GSs \cite{Barash}.

The stability of solitons is not sensitive to the value of the relative
Bragg reflectivity $\mu $, unlike the frustration $k$. In particular, the
variation of $\mu $ does not essentially change the shape of the stability
region displayed in Fig. \ref{semigaps}, nor the shape of the solitons shown
in Fig. \ref{solitons}.

\begin{figure}[tbp]
\renewcommand{\captionfont}{\small \sffamily} \renewcommand{\captionlabelfont}{}
\centering
\includegraphics[width=4in]{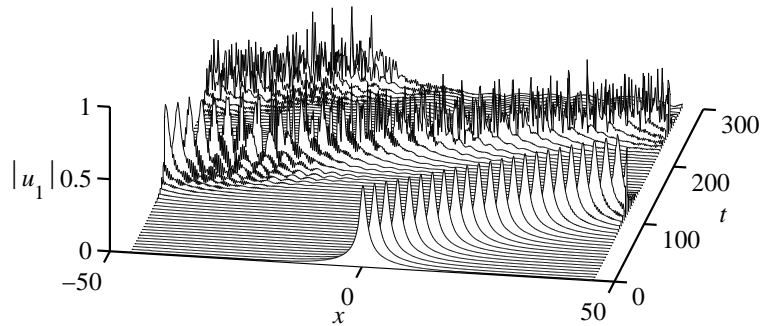} \renewcommand{\figurename}{Fig.}
\caption{A typical example of the destruction of an unstable
soliton, observed in direct simulations of Eqs.
(\protect\ref{tx}). Here (as well as in Figs. 4 and 5 below) shown
is the evolution of the field $\left\vert u_{1}\right\vert $; in
other components, the evolution is quite similar. The parameters
are $k=0,\protect\mu =0.5,c=0.4,\protect\omega =0.6$. The figure
uses the original coordinate $x$, rather than $\protect\xi $ from
Eq. (\protect\ref{c}).} \label{destruction}
\end{figure}

Note that stability of the solitons against non-oscillatory
perturbations (those corresponding to real stability eigenvalues)
is provided, at fixed $c$, by the above-mentioned VK criterion,
$dE/d\omega <0$ \cite{VK} [recall the soliton's energy $E$ is
defined as per Eq. (\ref{E})]. The applicability of the VK
criterion to the present model was not rigorously proven, but in
Ref. \cite{John} it was actually derived for the particular case
of $\mu =1$. Inspection of numerical data shows that all the
solitons satisfy this criterion; however, it does not preclude
instability against oscillatory perturbations (those corresponding
to pairs or quartets of complex instability eigenvalues), that is
why a part of the solitons are unstable. Further simulations
demonstrate that the development of the instability always leads
to complete destruction of the soliton, see a typical example in
Fig. \ref{destruction}.

\section{Collisions between moving solitons}

The next step in the study of the moving solitons is to consider
collisions between them, which is interesting in its own right,
and for various applications too \cite{Mak-moving}. As Eqs.
(\ref{tx}) are not integrable, an issue is whether collisions
between stable solitons will be elastic or not. We stress that
neither the stability of the solitons of the \textit{semi-gap}
type, considered in this paper, nor collisions between them were
considered in earlier works.

\begin{figure}[tbp]
\renewcommand{\captionfont}{\small \sffamily} \renewcommand{\captionlabelfont}{}
\centering
\includegraphics[width=4in]{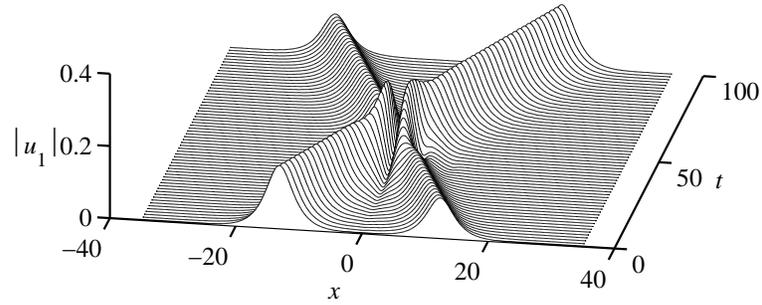} \renewcommand{\figurename}{Fig.}
\caption{A typical example of the elastic collision between two identical
stable solitons moving with the velocities $c_{1,2}=\pm 0.3$, the other
parameters being $k=1,\protect\mu =1,\protect\omega =2.08$.}
\label{elastic}
\end{figure}

\begin{figure}[tbp]
\renewcommand{\captionfont}{\small \sffamily} \renewcommand{\captionlabelfont}{}
\centering
\includegraphics[width=4in]{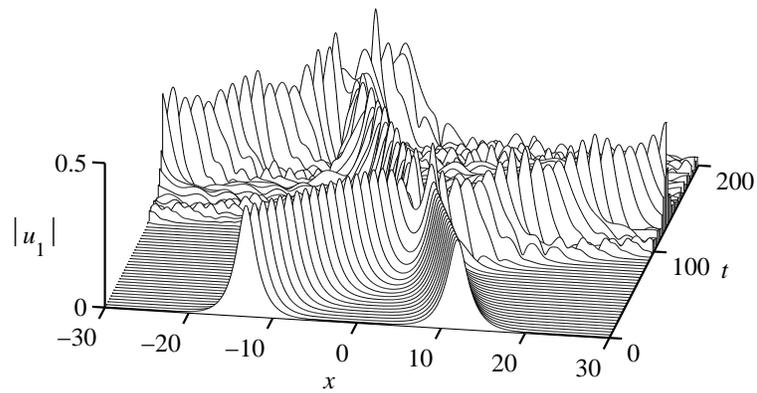} \renewcommand{\figurename}{Fig.}
\caption{A typical example of the inelastic collision between two stable
solitons moving with the velocities $c_{1}=-0.1$ and $c_{2}=+0.25$. The
other parameters are $k=0,\protect\mu =0.5,\protect\omega =1$.}
\label{inelastic}
\end{figure}

Exploring the stability region of the moving solitons (see Fig.
\ref{semigaps}), we have found that collisions between them may be
both completely elastic, leading to no other effect but finite
shifts of the colliding solitons, and strongly inelastic,
resulting in a partial destruction of the solitons, and generation
of large amounts of radiation. Typical examples of these outcomes
are shown in Figs. \ref{elastic} and \ref{inelastic}. The
interaction between the elastically colliding solitons has the
character of the bounce from each other.

The outcome of the collision strongly depends on the solitons'
velocities $c_{1,2}$. Therefore, the results can be naturally
summarized in the form of a region in the plane $\left(
c_{1},c_{2}\right) $ where the collisions are elastic; at the
border of the region, the transition between the two types of the
collision is found to be quite sharp. In turn, the size of the
region of the nondestructive (elastic) collisions depends on
proximity of the individual solitons to the instability border in
their parameter planes, see Fig. \ref{semigaps}.

\begin{figure}[tbp]
\renewcommand{\captionfont}{\small \sffamily} \renewcommand{\captionlabelfont}{} \centering\subfigure[]{\label{fig:subfig:a}
\includegraphics[width=3.2in]{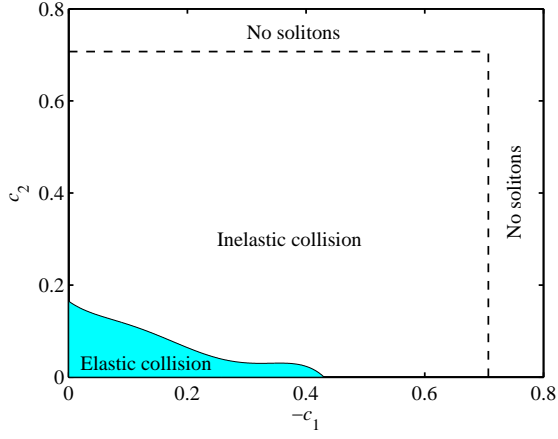}}
\subfigure[]{\label{fig:subfig:b}
\includegraphics[width=3.2in]{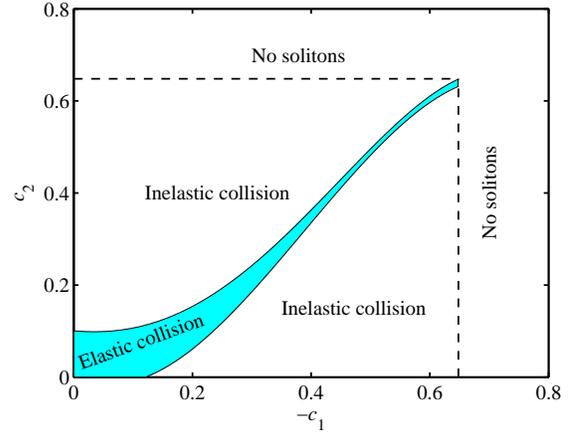}}
\subfigure[]{\label{fig:subfig:c}
\includegraphics[width=3.2in]{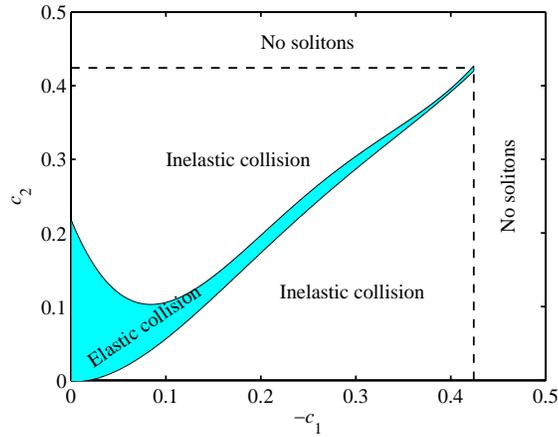}}
\par
\renewcommand{\figurename}{Fig.}
\caption{Areas of elastic and inelastic collisions between solitons in the
plane of their velocities, $\left( c_{1},c_{2}\right) $ (the values up to
the limiting value, $c_{\max }=1/\protect\sqrt{2}\approx \allowbreak 0.71$,
are shown). The fixed parameters are $k=0$ and $\protect\mu =0.5$. The
frequency is $\protect\omega =1.0$ (a), $1.2$ (b), $1.4$ (c). The horizontal
and vertical dashed lines limit the velocity intervals of the stable
solitons, as per Fig. \protect\ref{semigaps}.}
\label{cc}
\end{figure}

First, in Fig. \ref{cc} we present these results for the zero-frustration
case ($k=0$), when the stability region in Fig. \ref{semigaps}(a) is quite
large. As it follows from the inspection of Fig. \ref{semigaps}(a), the
difference between the three cases displayed in Fig. \ref{cc} is that in the
case of $\omega =1$ (a) the colliding solitons are close to the instability
border, while in the other cases, $\omega =1.2$ (b) and $\omega =1.4$ (c),
they are located deeper in the stability region (in the latter case, the
soliton is actually close to the edge of the semi-gap, i.e., the existence
border of the solitons). Accordingly, there is a notable difference in the
shape of the elasticity region in the $\left( c_{1},c_{2}\right) $ plane: in
the first case, the collisions may be elastic only if the velocities are
small, while in the other cases, the solitons with large but nearly opposite
velocities, $c_{1}\approx -c_{2}$, collide elastically even if the
velocities are large.

\begin{figure}[tbp]
\renewcommand{\captionfont}{\small \sffamily} \renewcommand{\captionlabelfont}{} \centering\subfigure[]{\label{fig:subfig:a}
\includegraphics[width=3.2in]{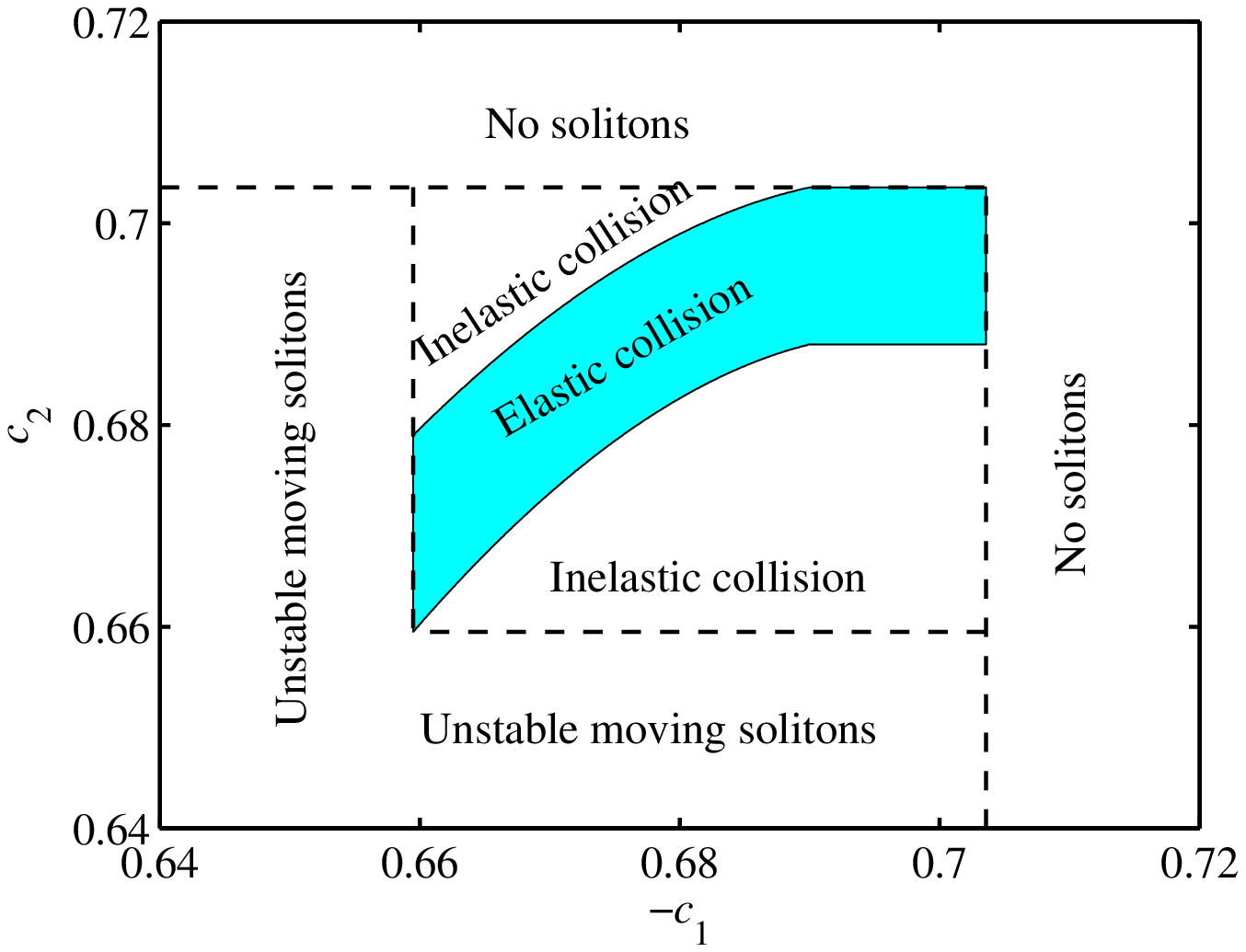}}
\subfigure[]{\label{fig:subfig:b}
\includegraphics[width=3.2in]{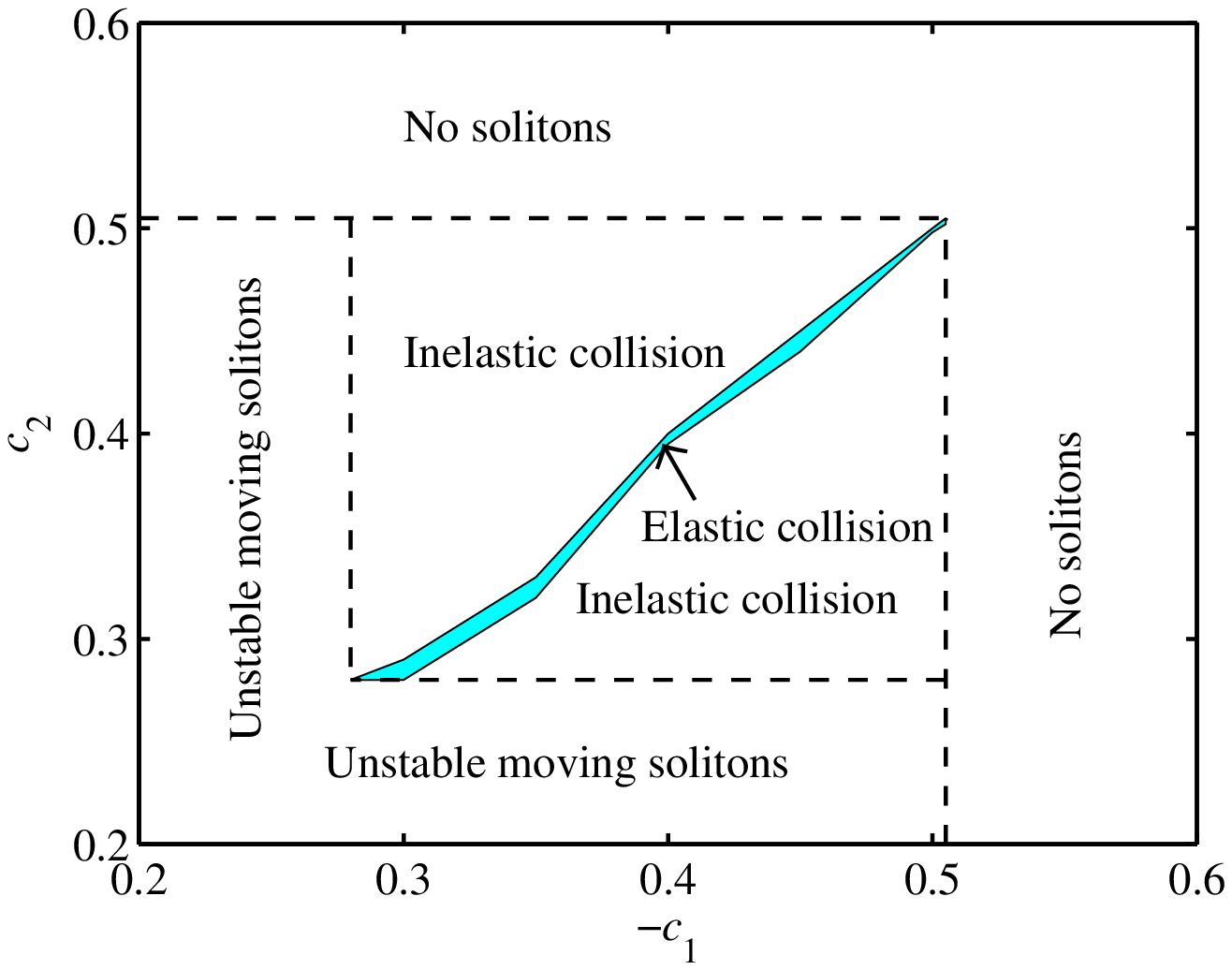}}
\subfigure[]{\label{fig:subfig:c}
\includegraphics[width=3.2in]{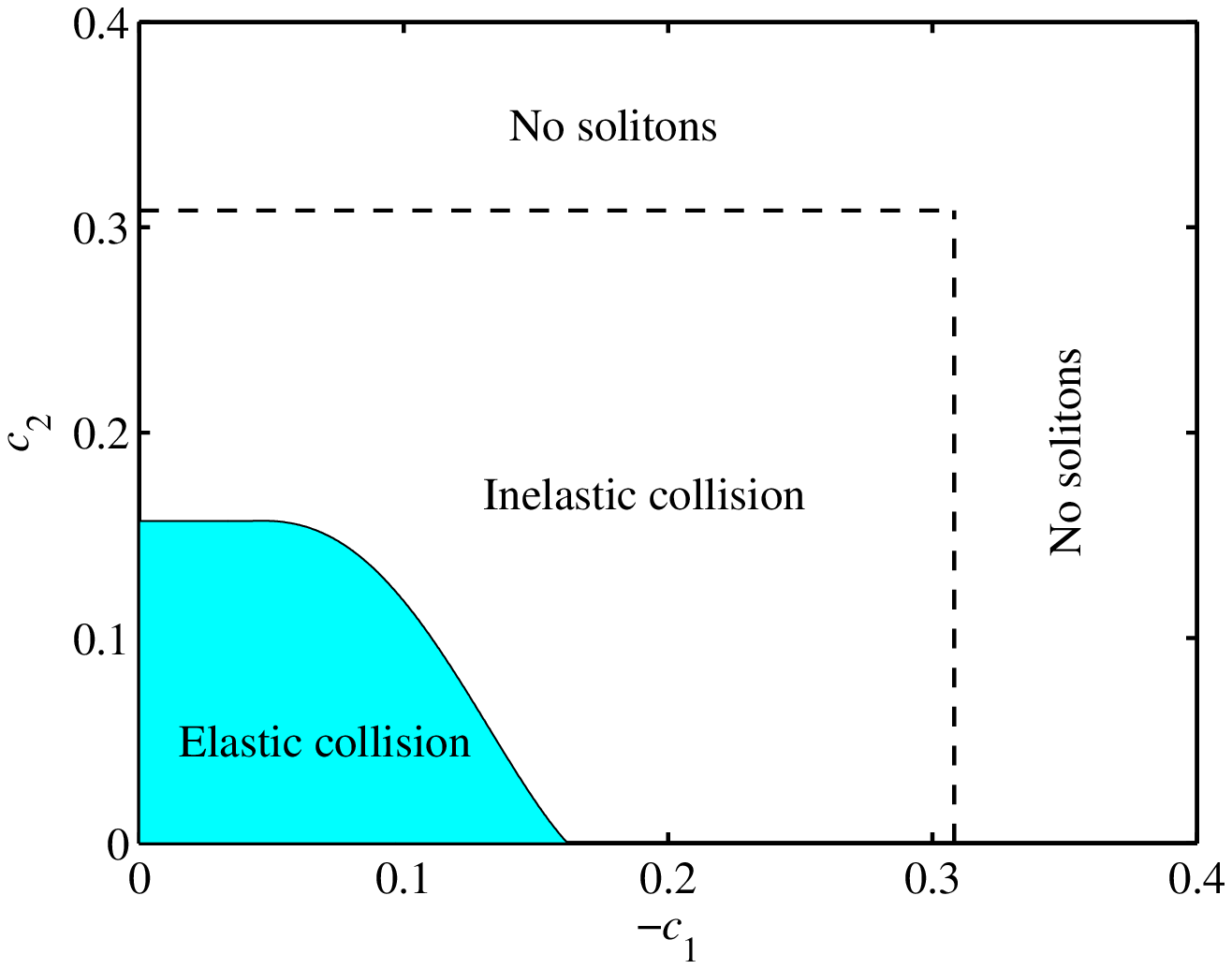}}
\par
\renewcommand{\figurename}{Fig.}
\caption{The same as in Fig. (\protect\ref{cc}) for the case of
nonzero frustration, $k=0.005$ and $\protect\mu =1$. The fixed
frequency is $\protect\omega =1.1$ (a), $1.7$ (b), $1.9$ (c).}
\label{cc2}
\end{figure}

As it was stressed above, in the case of finite frustration, the velocity
interval corresponding to the stable solitons may be limited not only from
above, but also from below. The collisions between unstable solitons being
definitely destructive, the latter fact strongly affects the region of
elastic (nondestructive) collisions in the $\left( c_{1},c_{2}\right) $
plane for $k\neq 0$, as is seen in Fig. (\ref{cc2}).

As it was mentioned above, in the currently feasible experiments with BG
solitons, their velocity cannot be made smaller than, approximately, half
the maximum group velocity, which is $1/\sqrt{2}$ in the present notation.
For this reason, it is quite important that the elasticity regions shown in
Figs. \ref{cc}(b,c) and \ref{cc2}(a,b) extend to large values of $c$.

Lastly, it is necessary to mention that, as well as the stability
of individual solitons, the outcome of the collisions is not
sensitive to the relative BG reflectivity $\mu $, on the contrary
to the strong dependence on the frustration $k$. In fact, the
elasticity regions show almost no dependence on $\mu $, therefore
the examples shown in Figs. \ref{cc} and \ref{cc2} are quite
generic ones.

\section{Conclusion}

In this paper, we have presented results of a systematic study of the
stability and collisions of moving four-wave solitons in the recently
introduced model of the Bragg cross-grating in planar optical waveguides
with the Kerr nonlinearity. We have focused on the case when the system's
linear spectrum does not contain any true bandgap, but rather features
semi-gaps, in which soliton families may be found. In recent works, only
zero-velocity solitons in this four-wave system have been studied. That case
is far from the current experiments, which generate solitons whose velocity
exceed, roughly, half the maximum group velocity. We have found the
semi-gaps for the moving solitons in the analytical form, and demonstrated
that numerically found solitons completely fill the semi-gaps. Stability of
the moving solitons was determined by direct simulations, revealing a
stability border which strongly depends on the frustration parameter. It is
interesting that a previously unknown situation is possible in the present
model, when the velocity interval for stable solitons is limited not only
from above, but also from below.

Collisions between stable solitons may be both elastic, so that they bounce
from each other, and strongly inelastic, leading to a partial destruction of
the solitons, with a sharp border between the two outcomes. A region of
elastic collisions was identified, with a well-pronounced peculiarity: if
the individual solitons are close to the stability border, they collide
elastically only if their velocities $c_{1}$ and $c_{2}$ are small; however,
if the solitons are more stable, the collision remains elastic in a strip
around the line $c_{1}=-c_{2}$. As well as the stability of individual
solitons, the elasticity region strongly depends on the frustration
parameter, but is insensitive to the relative reflectivity of the two Bragg
gratings which constitute the cross configuration.

\section*{Acknowledgements}

This work of B.A.M. was supported, in a part, by the Israel Science
Foundation through the grant No. 8006/03. T.M. and B.A.M. appreciate
hospitality of the Department of Electronics Engineering at the City
University of Hong Kong.


\begin{thebibliography}{99}
\bibitem{Kashyap} R. Kashyap, \textit{Fiber Bragg gratings }(Academic Press:
San Diego, 1999).

\bibitem{AcevesWab} A.B. Aceves and S. Wabnitz, Phys. Lett. A \textbf{141},
37 (1989); D. N. Christodoulides and R.I. Joseph, Phys. Rev. Lett.
\textbf{62}, 1746 (1989).

\bibitem{GSstability} B.A. Malomed and R. S. Tasgal, Phys. Rev. E
\textbf{49}, 5787 (1994)

\bibitem{Barash} I. V. Barashenkov, D. E. Pelinovsky, and E. V. Zemlyanaya,
Phys. Rev. Lett. \textbf{80}, 5117 (1998).

\bibitem{Trillo} A. De Rossi, C. Conti, and S. Trillo, Phys. Rev. Lett.
\textbf{81}, 85 (1998).

\bibitem{Mayer} J. Schollmann and A. P. Mayer, Phys. Rev. E \textbf{61},
5830 (2000).

\bibitem{MakDual} W. Mak, B.A. Malomed, and P. L. Chu, J. Opt. Soc. Am. B
\textbf{15}, 1685 (1998).

\bibitem{Alan} A. R. Champneys, B. A. Malomed, and M. J. Friedman, Phys.
Rev. Lett. \textbf{80}, 4169 (1998).

\bibitem{Tsoy} E. N. Tsoy and C. M. de Sterke, J. Opt. Soc. AM. B
\textbf{18}, 1 (2001).

\bibitem{Javid} J. Atai, B.A. Malomed, Phys. Rev. E \textbf{62}, 8713
(2000), \textit{ibid}. E \textbf{64}, 066617 (2001); Phys. Lett. A \textbf{\
298}, 140 (2002).

\bibitem{antisymm} G. Curatu, S. LaRochelle, C. Par\'{e}, and P. A.
Belanger, Electron. Lett. \textbf{38}, 307 (2002).

\bibitem{Georg} R. Grimshaw, B.A. Malomed, and G. A. Gottwald, Phys. Rev. E
\textbf{65}, 066606 (2002).

\bibitem{Chi} H. Y. Tseng and S. Chi, Phys. Rev. E \textbf{66}, 056606
(2002).

\bibitem{defect} R. H. Goodman, R. E. Slusher, and M. I. Weinstein, J. Opt.
Soc. Am. B \textbf{19}, 1635 (2002); W. C. K. Mak, B. A. Malomed, and P. L.
Chu, \textit{ibid}. \textbf{20}, 725 (2003); Phys. Rev. E \textbf{67},
026608 (2003).

\bibitem{experiment} B. J.~Eggleton, R. E.~Slusher, C. M.~de Sterke, P.
A.~Krug, and J. E.~Sipe, Phys. Rev. Lett. \textbf{76}, 1627 (1996); C. M. de
Sterke, B. J. Eggleton, and P. A. Krug, J. Lightwave Technol. \textbf{15},
1494 (1997); B. J. Eggleton, C. M. de Sterke, and R. E. Slusher, J. Opt.
Soc. Am. B \textbf{16}, 587 (1999).

\bibitem{OldReview} C. M. de Sterke and J. E. Sipe, Progr. Opt. \textbf{33},
203 (1994).

\bibitem{NewReview} C. Conti, G. Assanto, and S. Trillo, J. Nonlin. Opt.
Phys. Mat. \textbf{11}, 239 (2002).

\bibitem{BEC} R. G. Scott, A. M. Martin, T. M. Fromhold, S. Bujkiewicz, F.
W. Sheard, and M. Leadbeater, Phys. Rev. Lett. \textbf{90}, 110404 (2003);
B. Eiermann, T. Anker, M. Albiez, M. Taglieber, P. Treutlein, K. P. Marzlin,
and M. K. Oberthaler, Phys. Rev. Lett. \textbf{92}, 230401 (2004).

\bibitem{Feng} J. Feng, Opt. Lett. \textbf{18}, 1302 (1993).

\bibitem{IlyaTemporal} I. M. Merhasin and B. A. Malomed, J. Optics B: Quant.
Semiclass. Opt. \textbf{6}, S323 (2004).

\bibitem{IlyaSpatial} I. M. Merhasin and B. A. Malomed, Phys. Lett. A
\textbf{327}, 296 (2004).

\bibitem{John} N. Ak\"{o}zbek and S. John, Phys. Rev. E \textbf{57}, 2287
(1998).

\bibitem{VK} M.G. Vakhitov and A.A. Kolokolov, Radiophys. Quantum Electr.
\textbf{16}, 783 (1973); see also a review by L. Berg\'{e}, Phys. Rep.
\textbf{303}, 260 (1998).

\bibitem{Mak} W. C. K. Mak, B. A. Malomed, and P. L. Chu, Phys. Rev. E
\textbf{69}, 066610 (2004).

\bibitem{Mak-moving} W. C. K. Mak, B. A. Malomed, and P. L. Chu, Phys. Rev.
E \textbf{68}, 026609 (2003).
\end{thebibliography}
\end{document}